\documentclass[twocolumn,preprintnumbers,amsmath,amssymb]{revtex4}

\usepackage{graphicx}
\usepackage{dcolumn}
\usepackage{bm}

\begin{document}

\title{Dynamic detection of a single bacterium: nonlinear rotation rate shifts of driven magnetic microsphere stages}

\author{Brandon H. McNaughton$^{1,2}$}
\author{Rodney R. Agayan$^{1,2}$}%
\author{R. Kopelman$^{1,2}$}

 \email{kopelman@umich.edu}
 \homepage{http://www.umich.edu/~koplab}
\affiliation{%
$^1$Applied Physics Program, University of Michigan, Ann Arbor, MI
48109}
\affiliation{%
$^2$Department of Chemistry, University of Michigan, Ann Arbor, MI
48109-1055
\\}

\date{\today}

\begin{abstract}
We report on a new technique which was used to detect single
Escherichia coli that is based on the changes in the nonlinear
rotation of a magnetic microsphere driven by an external magnetic
field. The presence of one Escherichia Coli bacterium on the surface
of a 2.0 $\mu m$ magnetic microsphere caused an easily measurable
change in the drag of the system and, therefore, in the nonlinear
rotation rate. The straight-forward measurement uses standard
microscopy techniques and the observed average shift in the
nonlinear rotation rate changed by a factor of $\sim$3.8.
\end{abstract}

\maketitle


Magnetic microspheres and nanoparticles have been used for a variety
of medical applications and incorporated into various diagnostic
techniques \cite{haukanes:amb,olsvik1994mst,gu2003ubm}. While
magnetic particles have proven to be extremely useful, they have
been generally utilized in techniques that depend on the
translational properties of magnetic particles, such as magnetic
separation, giant magneto-resistive (GMR) sensors
\cite{rife2003dap}, and magnetic tunnel junctions (MTJ) sensors
\cite{shen2005sds}; however, it is possible through standard
microscopy techniques, to monitor the rotational behavior of single
magnetic particles or small chains of them
\cite{anker2003mmo,biswal2004mlc,lapointe2005sad,korneva2005cnl,mcnaughton2006jpcb}.
These small magnetic systems have been utilized to improve
immunoassays \cite{petkus2006dfc}, to act as micro-mixers
\cite{biswal2004mlc}, study microrheology
\cite{lapointe2005sad,behrend2005mmo} and even to reduce interfering
background in fluorescent spectroscopy measurements
\cite{anker2003mmo}. While single bacteria have been detected in
fluid using nanoparticles \cite{zhao2004rbs}, a dynamic application
that has yet to be developed is the detection of single
micro-biological agents using magnetic particles. We report on the
first measurement of a single bacterium using a rotating
micro-stage.

This new method, being development by the authors, is based on the
nonlinear rotation that a magnetic microsphere undergoes when driven
by a rotating magnetic field
\cite{mcnaughton2006jpcb,mcnaughton2006pms}. The effect occurs when
the viscous torque that arises from rotational drag is comparable to
the magnetic torque created by the external driving field. At low
external driving frequencies, the magnetic particle rotates
continuously and is synchronous with the external field, but at
sufficiently high external driving frequencies, the particle becomes
asynchronous with the driving field. The external driving frequency,
where the magnetic particle goes from linear to nonlinear
(synchronous to asynchronous) rotation, is dependent on
environmental conditions in addition to the particle properties and
is given by \cite{mcnaughton2006jpcb,mcnaughton2006pms}
\begin{equation}
\Omega_c = mB/\kappa\eta V,
\end{equation}
where $m$ is the magnetic moment, $B$ is the external magnetic
field, $\kappa$ is the shape factor, $\eta$ is the dynamic
viscosity, and $V$ is volume. The rotational dynamics of an actively
rotated magnetic particle are then given by
\begin{eqnarray}\label{freq_eqn}
\left<
 \dot{\theta}\left> = \left\{
\begin{array}{l l}
  \Omega & \quad \text{} \Omega<\Omega_c\\
  \Omega-\sqrt{\Omega^2-\Omega_c^2} & \quad \text{} \Omega>\Omega_c\\ \end{array}. \right.
\right.\right.
\end{eqnarray}
where $\left<
 \dot{\theta}\right>$ is the particle's average rotation rate and $ \Omega$
 is the driving frequency of an external magnetic field. Equation \ref{freq_eqn} holds for
 low Reynolds number environments ($Re\ll1$) and for our
 system $Re \approx 5\times 10^{-6}$.

\begin{figure} 
\begin{center}
  \includegraphics[width = 3 in]{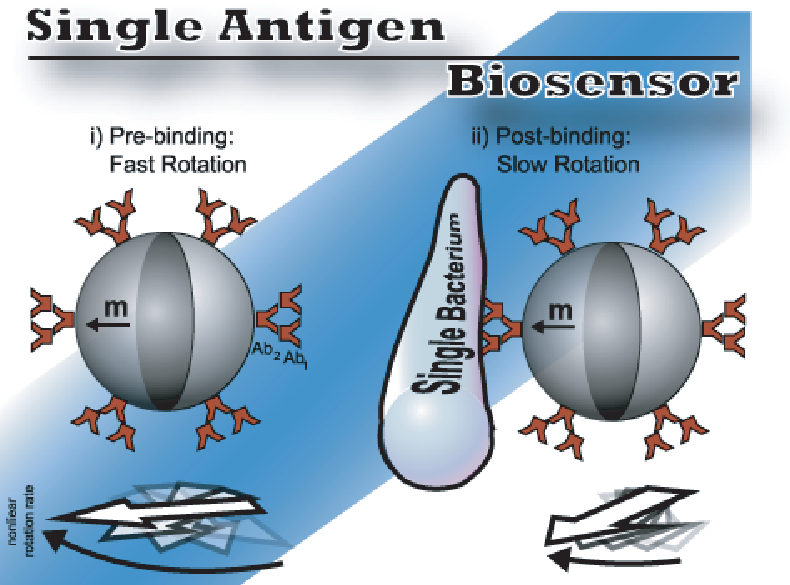}\\
  \caption{Schematic of the nonlinear rotation rate changes that a magnetic
microsphere undergoes when bound to a bacterium. The magnetic
microspheres are functionalized with a secondary (Ab2) and primary
(Ab1) antibody.}
  \label{figure1}
\end{center}
\end{figure}

Nonlinear rotation occurs when $\Omega>\Omega_c$ \cite
{shelton2005nmo} and we propose that this rate can be used to detect
single microbiological agents. The parameters that are important in
biological detection are shape and volume because of the drag
changes that occur when a bacteria binds to a microsphere. So, when
a bacteria attaches to a nonlinear rotating magnetic microsphere,
the micro-stage's volume and shape are drastically changed, which
produces more drag and, therefore, the rotation rate slows
considerably. This is shown schematically in Figure \ref{figure1}.
The technique is dynamic in the sense that a change in drag causes a
direct change in the nonlinear rotation rate. Past measurements have
shown that this technique can measure a change of drag caused by an
attachment of a 1.0 $\mu m$ particle to a 1.9 $\mu m$ nonlinear
rotating magnetic microsphere \cite{mcnaughton2006isd}.

The ability to develop sensitive diagnostic techniques using dynamic
systems has been a topic of high interest. One related technology
that has demonstrated extreme sensitivity in air and vacuum
environments is nanoelectromechanical systems (NEMS)
\cite{ekinci2005ns}. NEMS have been used to measure the mass of
single micro-biological agents like antibodies, viruses, and
bacteria \cite{ilic2000mri,ilic2001scd,ilic2004vdu}. One of the NEMS
detection schemes utilize the fact that the resonant oscillations
change when a micro-biological agent attaches, but the sensitivity
of such devices are drastically affected when operated in fluidic
environments \cite{vignola2006evl}. The idea underlying nonlinear
rotating magnetic particles can be used in a similar way, namely
when a biological agent attaches to the magnetic particle, the
nonlinear rotation frequency changes, but in such systems
sensitivity is unaffected, rather helped by drag. This allows for
single biological agent detection in fluidic environments. Thus, in
this letter we show that the nonlinear rotation frequency of single
magnetic microspheres on average rotates significantly slower when a
single bacterium is attached to their surface.


A 20 $\mu L$ aliquot of 2.0 $\mu m$ ferromagnetic microspheres
functionalized with goat antimouse IgG (Spherotech IL) was spread
onto a precut microscope slide and coated with $\sim$50 $nm$ of Al.
The sample was placed in a uniform magnetic field of ~1.4 $kOe$ so
that the magnetization would be perpendicular to the microscope
glass. The spheres were then rinsed with phosphate buffer solution
(PBS) at a pH of 7.2 and suspended in 500 $\mu L$ of PBS. The
suspended sample was centrifuged at 9000 $rpm$ for 8 $minutes$ and
resuspended in 500 $\mu L$ of PBS at a pH of 7.2. The sample was
centrifuged once more at 9000 $rpm$ for 8 $minutes$ and the
supernatant was removed. 100 $\mu L$ of mouse anti-E. Coli IgG
(Cortex Biochem, San Diego, CA) was added to the pellet of magnetic
microspheres. The primary antibody and the magnetic microspheres
were allowed to incubate at room temperature for 4 $hours$. The
excess primary antibody was removed by centrifuging the sample at
9000 $rpm$ for 8 $minutes$ and the supernatant was discarded.
Finally, the magnetic microspheres were resuspended into 500 $\mu L$
of PBS. At each of the above stages the sample was vortexed at 3000
$rpm$ for 15 seconds.

To make the bacteria fluorescent, a DsRed plasmid was used with
Escherichia Coli BL21(DE3) following previously described
transformation procedures \cite{shaner2004imr}. The bacteria were
allowed to reproduce until the sample had an optical density of .67
at 600 nm and was stored at 4 $^oC$. The magnetic microspheres with
the anti-E. Coli antibody were mixed 1:1 with the now fluorescent E.
Coli described.  To make binding more probable, the sample was
centrifuged at 9000 $rpm$ for 8 $minutes$. The sample was then
vortexed at 3000 $rpm$ for 15 $seconds$ and allowed to incubate. The
resulting sample had many single microspheres with 1-5 E. Coli bound
to their surface, where visual analysis was used to confirm the
presence of a single bacterium.

Two homemade $\sim$100 $\mu m$ thick fluidic cells were fabricated:
one fluidic cell contained the magnetic microsphere solution before
bacteria was added and the other had magnetic microspheres with
bacteria bound to their surfaces. Before being placed in the fluidic
cells, the samples were mixed with glycerol so that the
glycerol-water mass fraction was 0.5. The nonlinear rotation rates
for 20 single magnetic microspheres, without bacteria, were obtained
by monitoring the intensity fluctuations caused by light reflecting
off of the aluminum half-shell. From the other fluidic cell, 20
nonlinear rotation rates were obtained for single magnetic
microspheres with one E. Coli bound to their surfaces by monitoring
the intensity fluctuations caused by the bacteria fluorescence (for
more experimental and data analysis details see Refs
\cite{anker2003mmo,mcnaughton2006jpcb,mcnaughton2006pms}). The
rotation rates were then averaged and compared to determine the
rotation rate changes caused by the bacteria.

\begin{figure*} 
\begin{center}
  \includegraphics[width=6.5in]{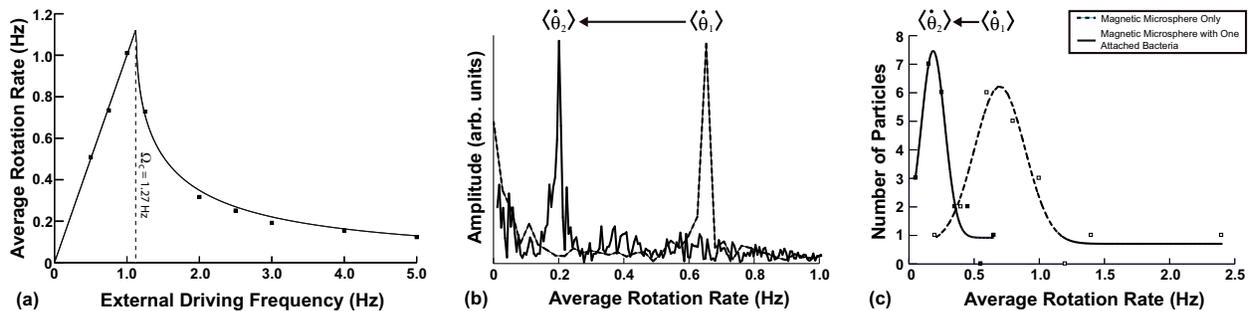}\\
  \caption{
a) The rotational response of a single magnetic particle with
attached bacteria at various external driving frequencies, where the
squares are data and the line is a theoretical fit. The average
nonlinear rotation rate b) for a typical particle with a bacteria
attached and for one without (dashed curve) and c) of 20 particles
in a fluidic cell with bacteria present and a fluidic cell without
bacteria. The magnetic microspheres with one bacterium attached
rotated $\sim$3.8 times slower then the blank micro-stages
(particles).}
  \label{figure2}
\end{center}
\end{figure*}

The theory for a single magnetic particle rotating in response to an
external driving field is well developed
\cite{mcnaughton2006jpcb,mcnaughton2006pms,cebers2006dam}, but
measurements have not previously been made for the case of a
magnetic particle attached to a bacteria. Figure \ref{figure2}(a)
shows the average rotation rate of such a system for increasing
external driving frequencies. The data is in good agreement with the
fit determined from Equation \ref{freq_eqn} and the critical
slipping rate, $\Omega_c$, was found to be 1.27 $Hz$. This
measurement shows that when a bacteria is bound to the surface of a
magnetic microsphere, the system can still be analyzed using
previously developed theory. Thus, a change in rotation rate can be
used to detect bacteria.

While the entire range of frequencies for magnetic particles with
and without bacteria could be scanned as was done in Figure
\ref{figure2}(a), it is much faster and more straight-forward to
only measure the value of the nonlinear rotation rate, $\left<
 \dot{\theta}\right>$, at a given
external driving frequency of $\Omega$. Figure \ref{figure2}(b)
shows this measurement for a typical magnetic microsphere with and
without a single bacteria attached to its surface. Figure
\ref{figure2}(c) shows the curves for the rotation rate of 20
particles in a fluidic cell with bacteria and for 20 particles in
one without bacteria. The presence of the bacteria on the surface of
the magnetic microspheres caused a measurable change in the average
rotation rate, namely the average rate of the particles at a driving
frequency of 4.0 $Hz$ changed from $\left<
 \dot{\theta_1}\right> = 0.72$ $Hz$ to $\left<
 \dot{\theta_2}\right> = 0.19$ $Hz$, a factor of $\sim$3.8. This
change in rotation rate is similar in value to our previous
measurements on a 1.0 $\mu m$ particle that was attached to a single
$\sim$1.9 $\mu m$ ferromagnetic microsphere
\cite{mcnaughton2006isd}. Once a bacteria is attached to a magnetic
microsphere, this technique could also be used to monitor single
bacteria growth, which could have significant application for the
study of single bacteria growth dynamics and in antibiotic
susceptibility measurements.


The ability to use the change in nonlinear rotation of magnetic
particles to detect bacteria has been demonstrated. The nonlinear
rotation rate of 2.0 $\mu m$ magnetic microspheres changed on
average from 0.72 $Hz$ without a bacterium to 0.19 Hz with a single
bacterium attached, where the driving oscillatory magnetic field was
at a frequency of 4.0 $Hz$.


The authors would like to thank Carol A. Fierke, Marcy Hernick, and
Tamiika K. Hurst for help with the bacteria growth and
transformations. Funding was provided by NSF-DMR $\#$ 0455330.
Related information, such as videos will be available online at
\textit{http://www.umich.edu/$\sim$koplab/moons.htm}.


\end{document}